\input amstex
\documentstyle{amsppt}
\loadmsbm

\topmatter

\title
Quantum Verification of Minimum Spanning Tree
\endtitle

\rightheadtext{Minimum Spanning Tree Verification}

\author
Mark Heiligman
\endauthor

\address
Intelligence Advanced Research Projects Activity,
Office of the Director of National Intelligence,
Washington, D.C.
\endaddress

\email
mark.i.heiligman\@ugov.gov
\endemail

\date
January 29, 2011
\enddate


\abstract 
Previous studies has shown that for a weighted undirected graph having
$n$ vertices and $m$ edges, a minimal weight spanning tree can be
found with $O^*\bigl(\sqrt{mn}\bigr)$ calls to the weight oracle.
The present note shows that a given spanning tree can be verified
to be a minimal weight spanning tree with only $O\bigl(n\bigr)$ calls
to the weight oracle and $O\bigl(n+\sqrt{m}\log n\bigr)$ total work. 
\endabstract

\thanks
{\it Disclaimer}. 
All statements of fact, opinion, or analysis expressed in this paper
are solely those of the author and do not necessarily reflect the
official positions or views of the Office of the Director of National
Intelligence (ODNI), the Intelligence Advanced Research Projects
Activity (IARPA), or any other government agency.  Nothing in the
content should be construed as asserting or implying U.S. Government
authentication of information or ODNI endorsement of the author's
views.
\endthanks

\endtopmatter

\document


\def\R{{\Bbb R}}

\def\frac#1#2{{{#1} \over {#2}}}

\def\ket#1{{\mid{#1}\rangle}}

\def\card#1{{\left\vert{#1}\right\vert}}

\def\th{{\hbox{\sevenrm th}}}

\head
Introduction
\endhead

\subhead
Problem Statement
\endsubhead

The determination of a minimal weight spanning tree of a weighted
undirected graph is a central problem in computational graph theory
and a number of well known classical algorithms address the problem
quite efficiently. This problem has also shown up in the realm of
quantum algorithms and the paper \cite{DHHM} provides nearly matching
upper and lower bounds for the problem. (The term ``nearly matching''
as used here means that the upper and lower bounds agree to withing a
power of the logarithm of the problem size.) The algorithm in
\cite{DHHM} uses some of the constructs that occur in the classical
minimal spanning tree algorithms, along with a somewhat sophisticated
version of the quantum minimum algorithm (which itself is based on
Grover's algorithm).

A closely related problem deals with the verification of minimal
spanning tree. In this formulation of the problem, both a weighted
graph and a spanning tree of that graph are given as inputs, and the
problem is to decide whether the given spanning tree is of minimal
weight (and if not to give a lower weight spanning tree). Based on
work of \cite{Ko}, a simple classical verification algorithm was given
in \cite{Ki}.

\proclaim{Problem}
Given a graph $G=(V,E)$ consisting of $n=\card{G}$ vertices and
$m=\card{E}$ edges along with a weight function on the edges
$w:\,E\rightarrow\R^+$, and a spanning tree $T=(V,F)$ with $F\subset
E$ and $\card{F}=n-1$, verify that $T$ is a minimal weight spanning
tree.
\endproclaim 

The goal of this paper is to develop a quantum algorithm for the
verification problem. We build heavily on the graph theory methods
given in \cite{Ki} and \cite{KPRS}.  Our quantum tool in this case is
a fairly simple version of Grover's algorithm. Nevertheless we are
able to show that verification is simpler than finding the solution
{\it ab initio}.

\subhead
Computational Models
\endsubhead

There is a basic question of how the graphs $G$ and $T$ are
presented, and this can critically affect the efficiency of the
algorithm. A graph can be presented by an adjacency matrix or by
a simple listing of its edges (and this may be either a sorted or
an unsorted list).

In the classical world, the problem statement is fairly simple. In the
quantum world, the graph is presented to the algorithm as an oracle,
and the complexity of the algorithm is measured in the number of
oracle calls necessary to solve the problem. 

Oracles can be applied in the classical world, as well, but they are
less indicative of the computational complexity of the problem than in
the quantum world. In the classical world, the entire graph needs to
be made available to the algorithm, so in the adjacency matrix model
there would be $O\bigl(n^2\bigr)$ calls to the oracle specifying the
graph, while in the edge list model there would be $m$ calls to the
oracle simply to get the entire graph into the computer.

There also has to be an oracle that gives the weight of an edge, and
in the adjacency matrix model or the edge list model, there would be
$m$ calls to the weight oracle.  It is useful to combine the graph
oracle with the weight oracle. In the adjacency matrix model, the
graph is extended to a complete graph and weight $+\infty$ is assigned
to all non-graph edges, with the oracle being given as a function
$w:\,V\times V\rightarrow\R^{+}\cup\{\infty\}$.
In the edge list model the oracle is a function, $e:\,[1,m]\rightarrow
V\times V\times\R^{+}$ where the first two components give the endpoints
of the $i^\th$ edge of the graph $G$ and the last component gives the
weight of that edge.

In this note, we consider both of these models, but from the quantum
perspective, the oracle has to be viewed as a reversible function that
then operates on quantum states. The two models to be considered here are:

(1) There is the weight oracle in the adjacency matrix model.  For
this model, a call to the quantum weight oracle is
$\ket{a,b,x}\rightarrow\ket{a,b,x\oplus w(a,b)}$ where $a,b\in V$ are
a pair of vertices. (Note that $x$ here is just some arbitrary initial
bit string.) For finding minimum weight spanning trees, this is bad if
the graph is moderately sparse. For checking the minimality of a
spanning tree, the input would consist of a simple listing of the
edges and would be of length $n-1$.

(2) There is the combined edge list and edge weight in the edge list
model.  For this model, a call to the quantum oracle is
$\ket{i,x,y,z}\rightarrow\ket{i,x\oplus a,y\oplus b,z\oplus w}$ where
$a,b\in V$ are a pair of vertices such that $(a,b)$ is the $i^\th$
edge of the graph $G$ and $w=w(a,b)$ is its weight. (Note that $x$,
$y$, and $z$ here is just some arbitrary initial bit strings.)

Note that the result for model (1) above will give an upper bound for
model (2), but both models will be considered in this note. Thus in
the adjacency matrix model, there will be a weight oracle given
$w:\,V\times V\rightarrow
\R^+\cup\{+\infty\}$ and the spanning tree to be checked for
minimality will be (classically) input as a list of edges
$T=\bigl\{(a_1,b_1),(a_2,b_2),\ldots,(a_{n-1},b_{n-1})\bigr\}$ with
$(a_i,b_i)\in V \times V$ for $i=1,\ldots,n-1$. We will also consider
the $e$ oracle in the edge list model. However, even there, the
spanning tree to be checked for minimality will still be classically
input as a list of edges, only now the spanning tree to be checked for
minimality will be (classically) input as a list of edge indices
$T=\bigl\{e_i,e_2\,\ldots e_{n-1}\bigr\}$ with $e_i\in[1,m]$.

In both of the above formulations, the subtree to be tested for
minimality by the quantum algorithm is input classically. This gives a
lower bound for the complexity of the quantum algorithm of $O(n)$,
since the algorithm has to at least read in all the (classical) input.
However, there are other possible statements of the problem.

(3) Given an oracle for the weights of $G$ (which is by default, also
an oracle for querying whether a given pair of points of $V$ is an
edge of $G$), the input could be by an oracle for the putative minimal
spanning tree. Thus, in the adjacency matrix model, there is a
function $mst: V\times V\rightarrow\{0,1\}$ where $mst(a,b)=1$ if
$(a,b)$ is an edge in $T$ and $mst(a,b)=0$ if $(a,b)$ is not an edge
in $T$, while in the edge list model,there is a function $mst: V\times
[1,m]\rightarrow\{0,1\}$ where $mst(i)=1$ if $i$ is an edge index in
$T$ and $mst(i)=0$ if $i$ is not an edge in $T$. In either case, the
problem then becomes to determine whether $mst$ is a correct
oracle. The complication is that there are now two oracles to count
calls to, and in principle there could be an operation curve of
tradeoffs.

In fact, this is almost certainly the case, because on one extreme the
minimal spanning tree can be found simply by computing it with a
quantum algorithm and then checking the $mst$ oracle for mismatches
with the minimal spanning tree found. This reduces to the problem:
Given a set $S$ and a subset $U\subset S$ and a (quantum) oracle
$p:\,S\rightarrow\{0,1\}$, is it the case that $p(s)=1$ if and only if
$s\in U$? Counting oracle calls here would seem to be a simple
application of Grover's algorithm.

\head
Minimal Weight Spanning Trees
\endhead

\subhead
Checking a Spanning Tree
\endsubhead

The key observation from \cite{Ki} is the following.  For a graph
$G=(V,E)$ and any spanning tree $T$ of $G$, there is a unique path
between any two edges $u,v\in V$.  $T$ is a minimal weight spanning
tree if and only if the weight of each edge $(u,v)\in E-T$ is greater
than or equal to the the heaviest edge in the path in $T$ between $u$
and $v$. What is needed is an easy way to find the weight of the
heaviest edge in the path in $T$ between $u$ and $v$.

The idea for checking a putative spanning tree $T$ for minimality is
to show that for any other edge of $G$ not in $T$, in the cycle formed
by including this edge, the highest weight edge in the cycle is
exactly this edge. There is no way that this edge can be part of a
minimum weight spanning tree.

This is to be checked for all edges of $G$, so by invoking Grover's
algorithm in the quantum setting, the total work is
$O\bigl(\sqrt{m}\bigr)$ times the work of checking an edge.  The
problem is that for checking an edge $(u,v)\in E-T$, the length of the
path in $T$ between $u$ and $v$ could be very large, perhaps even as
big as $n-1$, so even using Grover's algorithm to find the maximal
weight edge on this path is not adequately efficient.

\subhead
Boruvka Trees
\endsubhead

What is needed is a a new data structure that allows the maximal
weight edge on any path in $T$ to be found efficiently. The basic idea
for this comes from one of the earliest papers in computational graph
theory \cite{B}, that was the forerunner of several modern spanning tree
algorithms. The key properties of the Boruvka tree built from a
putative minimal spanning tree come from \cite{Ki} and will be 
summarized here without proof.

The idea of a Boruvka tree built from a spanning tree $T$ is that
$B_T$ is a tree whose leaves are the vertices $V$ and whose
internal nodes are to be viewed as subsets of $V$. 

In general for any graph $G=(V,E)$ and any spanning tree $T$ of $G$,
the Boruvka graph is a rooted tree of depth at most
$\lceil\log_2(\card{V})\rceil$. This Boruvka graph consists of
successively larger aggregations of elements of $V$. All nodes in a
Boruvka tree are subsets of $V$. The leaves are all the singleton sets
$\{v_i\}$ as $v_i$ runs over all the elements of $V$, and eventually
the root is formed, which will be $V$ itself. Any intermediate node in
a Boruvka tree is the union of its children.

A Boruvka tree is built from the bottom up. At each stage or level,
every node computes its nearest neighbor (i.e. the node that it is
closest to), and an edge is formed for all such nodes. The nodes of
the next level up are then the connected components of the graph of
the previous level. The weight of each branch is just the weight of
the edge of $G$ that was just added. The result is a rooted tree with
at most $2\,n$ nodes and $n$ leaves.

The key property of $B_T$ is that if $u,v\in V$ are a pair of vertices
and if $B_T(u,v)$ is the smallest subtree of $B_T$ that has both $u$
and $v$ as leaves, then the weight of the heaviest edge that connects
$u$ and $v$ in the original spanning tree $T$ is equal to the weight
of the heaviest edge in $B_T(u,v)$. The Boruvka tree $B_T$ is a full
branching tree, which means that it has a specified root, all its
leaves are at the same level, and each internal node has at least two
children.

The height of $B_T$ is at most $\lceil\log_2n\rceil$. Therefore once
$B_T$ has been constructed, finding the heaviest edge in $B_T(u,v)$
costs at most $O\bigl(\log n\bigr)$ operations. In fact, if $B_T$ has
already been built, then finding the heaviest edge in $B_T(u,v)$
requires no queries of the edge weight oracle.
Therefore, to check any edge in the original graph requires only one
oracle query, and total work at most $O\bigl(\log n\bigr)$.

\head
The Quantum Algorithm
\endhead

The Boruvka tree $B_T$ can be made with work $O\bigl(n\bigr)$, (not
just $O\bigl(n\log n\bigr)$ work), and can be done classically (see
\cite{Ki} and \cite{Ko}), the total number of oracle queries of the
weight function being $O\bigl(n\bigr)$, as well.  This is what makes this algorithm so
effective. Once the Boruvka tree of the input spanning tree has been
formed, it is possible to check whether the input spanning tree is
minimal.

To check any edge $(a,b)\in E$ from the original graph $G$, the
maximal weight of the edge in the path in $T$ that connects $a$ and
$b$ is easily found.  Simply start with the leaves $a$ and $b$ and go
up $B_T$ one level at a time until they meet at a common internal node
(which might be the root). Recording the maximal weight found in the
set of edges traversed in $B_T$ up the their common internal node
gives the maximal weight of the edge in the path that connects $a$ and
$b$. Since the height of $B_T$ is bounded by $\lceil\log_2n\rceil$,
the total work for this is $O\bigl(\log n\bigr)$, and no oracle calls
are required since $B_T$ is already built classically. To check if
$(a,b)$ is of lower weight than the maximal weight of the edge in the
path in $T$ that connects $a$ and $b$ only require one invocation of
the weight oracle. Of course, if this weight is less, then a lower
weight spanning tree than $T$ has been found by swapping out the
maximum weight edge in the path that connects $a$ and $b$ in $T$ with
the edge $(a,b)$.

Using Grover's algorithm over all vertex pairs $V \times V$ for 
the weight oracle, therefore requires $O\bigl(n\bigr)$ oracle queries, 
and $O\bigl(n\log n\bigr)$ total work. 

If an edge weight oracle is given, then it is possible to run Grover's algorithm 
over the original edge set $E$, which only requires  $O\bigl(\sqrt{m}\bigr)$ oracle queries, and
$O\bigl(\sqrt{m}\log n\bigr)$ total work. Since $m < n^2$, it follows
that the number of oracle queries for the quantum part of the
algorithm is less than the number of classical oracle queries needed
to construct $B_T$. Therefore the total number of oracle queries is
$O\bigl(n\bigr)$ and the total work is $O\bigl(n + \sqrt{m}\log n\bigr)$.

In conclusion, it is interesting that the verification of a putatively
correct answer can be accomplished with considerably less work than
that of finding the answer.

\enddocument

\end